\documentclass[12pt]{article}
\usepackage{amsmath}
\usepackage{amssymb}

\begin{document}

\centerline{\large \bf Relation between bulk
order-parameter correlation} \centerline{\large \bf function and
finite-size scaling}

\vspace*{0.5cm}
\centerline{X.S. Chen$^{1,2,a}$ and V. Dohm$^{1,b}$}

\centerline{$^1$ Institut f\"{u}r Theoretische Physik, Technische Hochschule
Aachen,}
\centerline{D-52056 Aachen, Germany}

\centerline{$^2$ Institute of Particle Physics, Hua-Zhong
Normal University,}

\centerline{Wuhan 430079, P.R. China}

\vspace*{0.5cm} \centerline{\it \today}


\begin{abstract}
We study the large-distance behavior of the bulk order-parameter
correlation function $G(\bf{r})$ for $T>T_c$ within the lattice
version of the $\varphi^4$ theory including lattice effects. We
also study the large-$L$ behavior of the susceptibility $\chi$ for
$T>T_c$ of the confined lattice system of linear size $L$ with
periodic boundary conditions. We find that the structure of the
large-$L$ behavior of $\chi$ of the confined system is closely
related to the structure of the large-distance behavior of
$G(\bf{r})$ of the bulk system. Explicit results are derived in
the spherical (large-$n$) limit and in one-loop order for general
dimensions $d>2$. For the lattice model with cubic symmetry we
find that finite-size scaling must be formulated in terms of the
anisotropic bulk correlation length (exponential correlation
length) that governs the exponential decay of $G(\bf{r})$ for
large $r$ rather than in terms of the ordinary isotropic bulk
correlation length $\xi$ defined via the second moment of
$G(\bf{r})$. We show that it is the exponential bulk correlation
length $\xi_1$ in the direction of the cubic axes that determines
the exponential finite-size scaling behavior of lattice systems in
a rectangular geometry. This result modifies a recent
interpretation concerning an apparent violation of finite-size
scaling in terms of the second-moment correlation length $\xi \neq
\xi_1$. Exact results for the one-dimensional Ising model
illustrate our conclusions. Furthermore we show for general $d>2$
that a description of finite-size effects for finite $n$ in the
entire region $0 \leq L/\xi \leq \infty$ requires $\it two$
different perturbative approaches that are applicable either to
the region $0 \leq L/\xi \lesssim O (1)$ or $O (1) \lesssim L/\xi
\leq \infty$, respectively. In particular we show that the
exponential finite-size behavior for $L/\xi\gg 1$ above $T_c$ is
not captured by the standard perturbation approach that separates
the homogeneous lowest mode from the inhomogeneous higher modes.
Consequences for the theory of finite-size effects above four
dimensions are discussed. We show that the two-variable
finite-size scaling form predicts an exponential approach $\propto
e^{-L/\xi_1}$ to the bulk critical behavior above $T_c$ whereas
the reduction to a single-variable scaling form implies a
power-law approach $\propto L^{-d}$.

\end{abstract}


$^a$ e-mail: chen@physik.rwth-aachen.de\newline
$^b$ e-mail: vdohm@physik.rwth-aachen.de


\newpage

\renewcommand{\theequation}{1.\arabic{equation}}
\setcounter{equation}{0}

\section{Introduction}

The fundamental length scale near a critical point is the bulk
correlation length $\xi$ which is a measure of the range of
correlations of the order-parameter fluctuations. In this paper we
consider lattice systems with cubic
symmetry at and above $T_c$. We assume short-range interactions.
The bulk order-parameter correlation function $G(\bf{r})$ between two
lattice points at relative separation $\bf{r}$ serves to define $\xi$ via the
second moment of $G(\bf{r})$ according to
\begin{equation}
\label{gleichung0}
\xi^2 = \sum_{\bf r} r^2 G({\bf r}) / \sum_{\bf r} G({\bf r})
\quad .
\end{equation}
In terms of
the Fourier transform $\hat{G}(\bf{k})$ this definition reads
\begin{equation}
\label{gleichung1} \xi^2 = \hat{G} ({\bf 0})
\frac{\partial}{\partial k^2} \; [ \hat{G}({\bf k}) ]^{-1}
\Big|_{k=0}\;.
\end{equation}
This is applicable to lattice
systems with cubic symmetry whose correlation function $\hat{G}
(\bf{k})$ has an isotropic small $\bf{k}$ behavior at $O(k^2)$.
The "second moment correlation length" (\ref{gleichung1}) is
widely used in field-theoretic calculations \cite{brezin-1976, amit-1984,
zinn-justin-1996}
since $\xi$ is identical with the inverse mass $m^{-1} = \xi$ that
enters the two-point vertex function $\Gamma^{(2)} ({\bf k}) =
\hat{G}({\bf k})^{-1}$.

The bulk correlation length (\ref{gleichung1}) plays a fundamental role
also in the formulation of the finite-size scaling behavior of
$\it{confined}$ systems. Consider, for example, the susceptibility
$\chi (t, L)\propto\hat{G}(\bf{0})$
of a ferromagnetic system for $t = (T-T_c) / T_c > 0$ in a finite
geometry with a characteristic size L and with periodic boundary
conditions. It is believed that the relative deviation from bulk critical behavior
$\chi (t, \infty) = A_\chi t  {^-}{^\gamma}$ has the asymptotic
(large $L$, small $t$) scaling form \cite{fisher-1971, brezin-1982, barber-1983,
privman-1990, binder-1997} below four dimensions

\begin{equation}
\label{gleichung2} \Delta\chi = \frac {\chi (t, \infty) - \chi (t,
L)}{ \chi (t, \infty)} = g (L / \xi)
\end{equation}

where $g(x)$ is a universal function in the entire range $0 \leq x \leq
\infty$. This means that any dependence of $\Delta\chi$ on the lattice constant $\tilde a$
is negligible in the asymptotic region $\xi \gg \tilde a$, $L \gg \tilde a$
for arbitrary ratios $L/\xi$, including the large-$L$ behavior at
fixed $\xi\gg \tilde a$ \cite{fisher-1971, gelfand-1990}.
In a recent paper we have shown \cite {chen-dohm-b10-1999}
that this statement is not valid in the regime $L/\xi \gg 1$.
Specifically, for cubic geometry of size $L \gg \xi \gg \tilde a$, we have found

\begin{equation}
\label{gleichung3} \Delta \chi = g (L/\xi)\exp
\left\{{\Gamma(\tilde{a}/\xi}) {\frac{L}{\xi}}\right\}\;.
\end{equation}

Here $g(x)$ is indeed universal but the $\tilde a$ dependent exponential
factor with the $\it nonuniversal$ function

\begin{equation}
\label{gleichung5} \Gamma (\tilde a/\xi) = \frac{1}{24}\; (\tilde
a/\xi)^2 + O \left[(\tilde a/\xi)^3\right]
\end{equation}

is non-negligible for $L\gtrsim 24\;\xi^3/\tilde a^2$, i.e., for
sufficiently large $L$ close to the bulk limit above $T_c$, even
in the asymptotic region $\xi \gg \tilde a$. Thus the
finite-size scaling form (\ref{gleichung2}) in terms of the
second-moment correlation length (\ref{gleichung1}) is not valid
for lattice systems, even below four dimensions. This conclusion
is based on one-loop results as well as on exact results in the
spherical limit of the $\varphi ^4$ lattice model at finite
lattice spacing \cite{chen-dohm-b10-1999}. We note that the
possibility of a non-negligible dependence of finite-size effects
on the lattice spacing $\tilde a$, even for $d<4$, was already
mentioned by Privman and Fisher \cite{privman-fisher-1984}. An as
yet unexplained $\tilde a$-dependence that is not consistent with
a finite-size scaling ansatz of the kind (\ref{gleichung2}) was
also found by Gelfand and Fisher \cite{gelfand-2} in the
interfacial free energy of the two-dimensional Ising model for $T
< T_c$.

No intuitive reasoning was given in our recent work \cite{chen-dohm-b10-1999} as to what
might be the physical origin for this unexpected failure of the finite-size
scaling property. It is the purpose of the present paper to
elucidate this unsatisfactory situation by further analyzing the
role played by the correlation length in both bulk and confined
lattice systems. Instead of the
second-moment correlation length (\ref{gleichung1}) we consider the "exponential
correlation length" $\xi _{\bf e}$ that governs the large-distance
behavior of the bulk system at fixed $T$ above $T_c$
\cite{fisher-1967},

\begin{equation}
\label{gleichung4} G (r {\mathbf{e}}) \sim B_d\;  r ^{(1-d)
/2}\exp (- r / \xi_{\mathbf e})\;,
\end{equation}

where ${\bf e} = {\bf r} / r$ is the unit vector in the direction of
$\bf r$ and where $\xi_{\bf e}$ is defined by

\begin{equation}
\label{gleichung5a} \xi ^{-1}_{\bf e} = -\lim_{r\to\infty} \{ [\ln G
(r {\bf e}) ] / r\}\; .
\end{equation}

We note that (\ref{gleichung4}) is expected to hold even well above
$T_c$ outside the asymptotic critical region.
For lattice systems, $\xi _{\bf e}$ is an anisotropic quantity. It
is expected, however, that the asymptotic ratio
$\;\lim_{t\rightarrow 0 +}\xi_{\mathbf e}/\xi\;$ becomes isotropic
and has a universal value \cite{tarko-1975, gottlob-1993, campostrini-1998,
provero-1998, caselle}, in agreement with our results.

In the present paper we shall call attention to the fact that the
non-universal
difference between $\xi$ and $\xi_{\mathbf e}$ is non-negligible
even asymptotically close to $T_c$ if the scaling function has an exponential form.
We shall show that lattice effects cause additive non-universal corrections to the
asymptotic form of $\xi_{\mathbf e}$ [see (\ref{gleichung7}) below].
Since $\xi_{\mathbf e}$ appears in the
exponent of the scaling functions these $\it additive$ corrections
turn into non-negligible $\it multiplicative$ overall factors for
the exponential scaling functions of both $G({\bf r})$ for $r\gg
\xi_{\bf e}$ and $\chi(t,L)$ for $L\gg\xi_{\mathbf e}$.

Because of the anisotropy of $\xi_{\bf e}$ for
lattice systems, there exist infinitely many $\xi_{\bf e}$ in
contrast to the unique isotropic quantity $\xi$. Since $\chi (t,L)
\propto \hat {G} ({\bf 0}) \propto \sum _{\bf r}  G ({\bf r})$ involves
all directions of $\bf r$ it is not clear a priori whether a
certain average of $\xi_{\bf e}$ or $\xi_{\bf e}$ in a particular
direction $\bf e$ enters the finite-size scaling form of $\chi (t,
L)$. Here we shall find that for cubic geometry the exponential finite-size effect is
determined by the correlation length $\xi_1 \equiv \xi^{}_{{\bf
e}_{1}}$ (and not by $\xi\neq\xi_1$) where ${\bf e}_1$ is the unit
vector in the direction of one of the cubic axes. Specifically we
derive an explicit relation between $\xi_1$ and $\xi$ in the
spherical limit (Sect. 3) and in one-loop order of the $\varphi^4$ lattice
model (Sect. 4). This relation reads at finite lattice spacing $\tilde a$

\begin{equation}
\label{gleichung6} \xi = \frac{\tilde{a}}{2} \left[ \sinh \left(\frac
{\tilde a} {2\xi_1}\right)\right]^{-1}
\end{equation}

\begin{equation}
\label{gleichung7} = \xi_1 \left[1 - \frac{1}{24}
\left(\frac {\tilde{a}}{\xi_1}\right)^2 + ... \right]\;.
\end{equation}

It is remarkable that the non-universal $\tilde a$ dependence of
(\ref{gleichung3}) is completely absorbed by the exponential correlation length
$\xi_1$ if the function $\xi (\xi_1)$ of (\ref{gleichung6}) is substituted into
the right-hand side of (\ref{gleichung3}). In the present paper we
shall show that for $L \gg \xi \gg \tilde a$

\begin{equation}
\label{gleichung8} g \Big(L/\xi (\xi_1)\Big)\exp \left\{\Gamma \Big(\tilde a/\xi(\xi_1)\Big) \frac{L} {\xi(\xi_1)} \right\}
 = g (L/\xi_1)\; ,
\end{equation}

hence universal finite-size scaling
below four dimensions $\it{at}\;{finite}\;{lattice}\; {spacing}$ is restored in the form

\begin{equation}
\label{gleichung9} \Delta\chi = g (L/\xi_1)
\end{equation}

with the universal function $g(x)$.

Thus, at one-loop order and in the spherical (large-$n$) limit, there is no
violation of finite-size scaling at finite lattice spacing below
four dimensions
provided that the exponential correlation length $\xi_1$ rather than $\xi$ is employed as the bulk
reference length.
We note that this result is not a general consequence of the
renormalizability of the $\varphi^4$ theory but is only an explicit
computational result for cubic (and rectangular) geometry in
one-loop order and in the spherical (large-$n$) limit. It is also
valid for the one-dimensional Ising model (Sect. 6). At
present it is an open question whether this result remains valid
at finite $n$ beyond one-loop order and whether it can be based on more general
arguments.
The renormalization-group (RG) arguments in the
limit $\tilde a \rightarrow 0$ \cite{brezin-1982} are not
sufficient to establish the complete finite-size scaling form for
lattice systems. Although these RG arguments for $\tilde a
\rightarrow 0$ lead to
the same universal scaling function $g(x)$ as our analysis at finite $\tilde a$,
they do not identify the $\tilde a$ dependent
finite-size scaling variable $x = L/\xi_1$.

We shall also show (Sect. 4) that $\it two$ different perturbative
treatments are necessary to describe the finite-size effects in
the entire asymptotic region $0 \leq L/\xi \leq \infty$ and that
the previous finite-size perturbation approach below four
dimensions \cite{brezin-1985, rudnick-1985, esser-1995} does not
capture the exponential structure of the scaling function $g(x)
\propto e^{-x}$ for $x \gg 1$.

The necessity of employing $\xi_1$ rather than $\xi$ is not
restricted to $d < 4$ dimensions. In Sect. 5 we shall discuss the
consequences of our results for lattice systems with $d > 4$ where
$\xi_1$ rather than $\xi$ should be incorporated in the
two-variable finite-size scaling form
\cite{chen-dohm-c9-1073-1998, privman-1983}. The inhomogeneous
modes are shown to yield $\Delta\chi \propto \;e^{-L/\xi_1}$ for
$L \gg \xi$ whereas the lowest-mode approximation
\cite{brezin-1985} and the corresponding single-variable
finite-size scaling form \cite{binder-1985} imply $\Delta\chi
\propto \; L^{-d}$.

Very recently an exponential size dependence has been confirmed by
Stauffer \cite{stauffer} in Monte Carlo simulations for the
magnetization of the Ising model in two, three and five dimensions
\cite{chen-dohm-stauffer}.

In summary, even though we will confirm through
(\ref{gleichung7}) (within our approximations)
that asymptotically close to the critical point the two
correlation lengths $\xi$ and $\xi_1$ are the same and
isotropic, the scaling form $g(x)$ of the leading finite-size
effect near the bulk limit is sensitive to the choice of
the correlation length because of the exponential dependence
of $g(x) \propto e^{-x}$ on the correlation length.

\newpage

\renewcommand{\theequation}{2.\arabic{equation}}
\setcounter{equation}{0}

\section{Lattice effect on the large-distance behavior of the bulk
order-parameter correlation function}

We consider a $\varphi^4$ lattice Hamiltonian for the variables
$\varphi_i$ on the lattice points ${\bf x}_i$ of a simple-cubic
lattice with the lattice constant $\tilde a$. The variables
$\varphi_i$ have $n$ components $\varphi_{i\alpha}$ with $\alpha =
1,2,...,n$ which vary in the range $-\infty\leq \varphi_{i\alpha}
\leq \infty$. We assume the statistical weight $\propto e^{-H}$
with

\begin{equation}
\label{gleichung10} H = \tilde a ^d \left\{\sum_i \left[\frac{r_0}
{2} \varphi ^2 _i + u_0 (\varphi^2 _i) ^2\right]+ \sum _{i,j}
\frac {1} {2 \tilde a ^2} J_{ij} (\varphi_i - \varphi_j)^2 \right
\}
\end{equation}

where $J_{ij}$ are the dimensionless couplings of a short-rang
interaction with cubic symmetry and where $k_B\; T \thickapprox
k_B\;T_c$ is absorbed in $r_0, u_0$ and $J_{ij}$. The variables
$\varphi_i$ have the dimension $\left[ \tilde a ^{(2-d) / 2} \right]$.
We are interested in the large-distance behavior of the bulk correlation
function

\begin{equation}
\label{gleichung11} G ({\bf{x}} _j - {\bf{x}} _0) = \frac {1} {n}
\langle{\varphi_j} \; {\varphi_0}\rangle = \int\limits_{\bf k}
\hat G ({\bf k}) e^{i {\bf k}({\bf x}_j - {\bf x}_0)}
\end{equation}

above $T_c$, normalized to the number of components $n$,
where

\begin{equation}
\label{gleichung12} \hat{G} ({\bf k}) = \frac{\tilde a ^d}{n}\sum
_{j} \langle{\varphi_j} \; {\varphi_0}\rangle e^{-i {\bf k} ({\bf
x}_j - {\bf x}_0)}
\end{equation}

with some fixed lattice point ${\bf x} _0$. In
(\ref{gleichung11}) $\int_{\bf k}$ stands for $(2 \pi)^{-d} \; \int d
{^d}k$ with a finite lattice-cutoff $\mid{k_m}\mid \leq
\pi/\tilde{a}, m = 1, 2, \ldots, d$. First we consider the
limit $n \rightarrow \infty$ at fixed $u_{0}n$ for $d > 2$ in which case we
obtain \cite{chen-dohm-b5-1998}

\begin{equation}
\label{gleichung13} \hat{G} ({\bf k}) ^{-1} = \hat{G} ({\bf 0})
^{-1} + \hat{J}_{\bf k}\;,
\end{equation}

\begin{equation}
\label{gleichung14} \hat{J}_{\bf k} = \frac {2} {{\tilde a} ^{2}}
[J ({\bf 0}) - J ({\bf k}) ] ,
\end{equation}

\begin{equation}
\label{gleichung15} J ({\bf k}) = ({\tilde a} / L)^d \sum _{i, j}
J {_i} {_j}\; e ^{-i {\bf k}({\bf x}_j - {\bf x} _i)}\; .
\end{equation}

$\hat G ({\bf 0})$ is determined by an implicit equation
\cite{chen-dohm-b5-1998}
which, however, will not be needed in the following since $\hat G
({\bf 0})$ can be expressed directly in terms of $\xi^2$. Using
the second-moment definition for the bulk correlation length $\xi$
according to (\ref{gleichung1}) we have \cite{chen-dohm-b5-1998}

\begin{equation}
\label{gleichung16} \hat G ({\bf 0}) = J_0 ^{-1} \xi^2\;,
\end{equation}

\begin{equation}
\label{gleichung17} J_0 = \frac {1}{d} ({\tilde a} / L)^d \sum
_{i, j} (J_{ij} / {\tilde a} ^2) ({\bf x}_i - {\bf x}_j)^2.
\end{equation}

For simplicity we consider a nearest-neighbor interaction $J > 0$
which yields

\begin{equation}
\label{gleichung18} J ({\bf k}) = 2 J \sum ^d _{m=1} \cos (k_m
\tilde a),
\end{equation}

\begin{equation}
\label{gleichung19} \hat J _{\bf k} = \frac {4J} {{\tilde a} ^{2}}
\sum ^d _{m=1} [1 - \cos (k_m \tilde a) ] = J_0 {\bf k} ^2 + O(k^2_i k^2_j)
\end{equation}

with $J_0 = 2 J$.

In summary we need to calculate the large-distance behavior of

\begin{equation}
\label{gleichung22} G ({\bf x}) = \frac {{\tilde a} ^{2}}
{2J}\int\limits_{\bf k}  e\;^{i\bf {k}{x}} \left\{ (\tilde {a}
/{\xi})^{2} + 2 \sum ^d _{m=1} [1 - \cos (k_m \tilde a) ]
\right\}^{-1}
\end{equation}

where we have chosen ${\bf x}_0 = {\bf 0}$ and ${\bf x}_j \equiv {\bf x}
= (x_1, x_2, ..., x_d)$ with Cartesian coordinates $x_m$. Eq.
(\ref{gleichung22}) is valid not only in the large-$n$ limit but
also in an ordinary perturbation calculation to one-loop order for
general $n$. (For the latter case a renormalization-group
treatment is carried out in Appendix B for $d < 4$.) A
representation of $G({\bf x})$ in terms of Bessel
functions of integer order $\nu$ (see, e.g., 9.6.19 of Ref.
\cite{abramowitz-1972})

\begin{equation}
\label{gleichung23} I_\nu (z) = \frac {1} {\pi} \int\limits ^\pi
_0 d\theta \; e ^{z \; \cos \; \theta} \cos (\nu \theta)
\end{equation}

can be given as

\begin{equation}
\label{gleichung24} G ({\bf x}) = \frac {{\tilde a} ^{2}} {2J}
\int\limits ^\infty_0 ds \;e ^{-(\tilde a / \xi) ^2s} \int\limits
_{\bf k} \;e \; ^{i\bf {k} {x}}\; \exp \left\{ - 2 s \sum ^d
_{m=1} {[1 - \cos (k_m \tilde a) ]} \right\}
\end{equation}

\begin{equation}
\label{gleichung25} = \frac {{\tilde a} ^{2-d}} {2J} \int\limits
^\infty_0 ds \; e ^{-(\tilde a / \xi) ^{2}s}\; e ^{-2ds}
\prod\limits ^{d} _{m=1}\; I_{\nu_{m}} (2s)
\end{equation}

with the integers $\nu_m = x_m / \tilde a$ . In general, $G (\bf
x)$ is an anisotropic function whose exponential large-distance
behavior $\sim \exp {\left(-|{\bf x}|/\xi_{\bf{e}}\right)}$ leads to the
definition of an anisotropic correlation
length $\xi_{\bf e} \neq \xi$ in the direction of the unit vector
$\bf e = \bf x / |\bf x|$ \cite{fisher-1967}. An explicit demonstration of the anisotropy of $\xi _{\bf
e}$ is given in Appendix A where the angular dependence of
$\xi_{\bf e}$ is calculated for the case
where $\bf e$ lies in the
2-dimensional $x_{1} - x_{2}$ plane of the d-dimensional bulk
system. For our present purpose it suffices to consider only the
special case where ${\bf {e}} = {\bf {e}}_1 = (1,0,0 ...)$ is the
unit vector along one of the cubic axes. Then we have ${\bf x} =
(x, 0, 0 ...)$ and ${\bf kx} = k_{x}x \; .$ The corresponding
correlation function is denoted by $C (x) = G (\bf x)$ which is
obtained from (\ref{gleichung25}) as

\begin{equation}
\label{gleichung26} C (x) = \frac {{\tilde a} ^{2-d}} {2J}
\int\limits ^\infty_0 ds \; e ^{-(\tilde a / \xi) ^{2}s}\; e
^{-2ds} [I_0 (2s)]^{d-1} I_{x/\tilde a} (2s) \;.
\end{equation}

This result is valid for arbitrary $x/\tilde a$ and $\xi/\tilde a$
and therefore does not yet have a scaling form.
In Appendix A the large-$|x|$ behavior of $C(x)$ at arbitrary fixed $\xi/\tilde a$
is derived. The result is

\begin{equation}
\label{gleichung27} C (x)\; = \;  \frac {{\tilde a} ^{2-d}}
{4J} \left(\frac {\tilde a} {2\pi|x|}\right)^{(d-1)/2} \left[\sinh \left(\frac {\tilde
a}{\xi_1}\right)\right] ^{(d-3)/2} \; e \; ^{-|x|/\xi_1} \left[1 +
O (|x|^{-1})\right]\;.
\end{equation}

We see that in the large-$|x|$ limit the natural reference length is the
exponential correlation length $\xi_1 \equiv \xi_{{\bf e}_{1}}$ in the
direction of one of the cubic axes
rather than the second-moment correlation length $\xi$. The exact
relation between $\xi_1$ and $\xi$ for $n \rightarrow\infty$
reads

\begin{equation}
\label{gleichung28} \xi ^{-1} = \frac {2} {\tilde a} \sinh
\left(\frac {\tilde a} {2\xi_1}\right)
\end{equation}

or

\begin{equation}
\label{gleichung29} \xi ^{-1} _{1} = \frac {2} {\tilde a}
{\rm{arsinh}}
\left(\frac {\tilde a} {2\xi}\right)
\end{equation}

The difference between $\xi$ and $\xi_1$ is a true lattice effect
that disappears in the formal limit $\tilde {a} \rightarrow
0$. Eqs. (\ref{gleichung28}) and (\ref{gleichung29}) are also
valid for finite $n$ in one-loop order above $T_c$ but in two-loop
order and beyond we expect (small) corrections to
(\ref{gleichung28}) and (\ref{gleichung29}) for finite $n$.

Eqs.(\ref{gleichung27})-(\ref{gleichung29}) are valid
for arbitrary $\xi_1/\tilde a$ even well above $T_c$. In the asymptotic
region $\xi_1 \gg \tilde a$,
(\ref{gleichung27}) attains the scaling form

\begin{equation}
\label{gleichung29b} C (x) \sim \left(\tilde a /|x|
\right)^{d-2+\eta}\;\Phi \left(|x| / \xi_1 \right)
\end{equation}

with the scaling function for $|x|/\xi_1\gg 1$

\begin{equation}
\label{gleichung29c} \Phi \left(|x|/\xi_1\right)\;=\;\tilde A\; \frac{{\tilde a}^{2-d}}{{4J(2\pi)}^{(d-1)/2}}
\left(\frac{|x|}{\xi_1} \right)^{\frac{1}{2}(d-3)+\eta} \exp \left(-|x|/\xi_1\right)
\end{equation}

where $\eta=0$ and $\tilde A = 1$ in the present case of the limit
$n \rightarrow \infty$. Eqs. (\ref{gleichung29b}) and
(\ref{gleichung29c}) are also valid for general $n$ in one-loop
order for $d < 4$, see Appendix B. In this case we have a critical
exponent $\eta > 0$ and an amplitude $\tilde A \neq 1$ which we
obtain from a RG treatment at finite $\tilde a$
\cite{chen-dohm-b10-1999}, applied to the bare one-loop result
(\ref{gleichung22}) for general $n$, as described in Appendix B.
Eqs. (\ref{gleichung29b}) and (\ref{gleichung29c}) are also valid
in one-loop order for general $n$ and $d > 4$ where $\eta = 0$ and
$\tilde A = 1$, apart from $O$ $(u^{2}_{0})$ corrections.

Close to $T_c$ where
both $\xi$ and $\xi_1$ diverge, an expansion of
(\ref{gleichung29}) yields

\begin{equation}
\label{gleichung30} \xi_1 = \xi\left [1 + \frac {1}
{24} \left(\frac {\tilde a} {\xi}\right) ^2 + ... \right]
\end{equation}

Thus, for $n \rightarrow\infty$, $\xi_1$ and $\xi$ become identical sufficiently close to
$T_c$. (This is also valid for finite $n$ in one-loop
order above $T_c$ but in two-loop order and beyond
the asymptotic value of the ratio $\xi_1/\xi$ for $T \rightarrow T_c$
is expected to become different from 1, see e.g. Refs.\cite{tarko-1975, gottlob-1993,
campostrini-1998, provero-1998, caselle}.)
Therefore $\xi_1$ can
be replaced by $\xi$ in the prefactor of
(\ref{gleichung29c}). We emphasize, however, that a replacement
of $\xi_1$ by $\xi$ is not possible in the exponential part of
(\ref{gleichung27}) and (\ref{gleichung29c}), even arbitrarily close to $T_c$. This is
seen by substituting (\ref{gleichung30}) into the exponential function of
(\ref{gleichung27}) and (\ref{gleichung29c}),

\begin{equation}
\label{gleichung31} e^{-|x|/\xi_1} = \exp\left(|x|\tilde a^2 / 24
\xi^3\right)
e ^{-|x|/\xi} \;,
\end{equation}

\begin{equation}
\label{gleichung31a} C (x) \sim (\tilde a/|x|)^{d-2+\eta}\;\Phi\;
(|x|/\xi)\; \exp \left (|x| \tilde a^2 / 24\xi^3 \right )\;.
\end{equation}

Now the $\it{additive}$ correction in (\ref{gleichung30}) has turned
into an exponential non-universal $\it{prefactor}$ in
(\ref{gleichung31}) and (\ref{gleichung31a})
that cannot be simply replaced by 1 and that is by no means
negligible for sufficiently large $|x|\gtrsim 24 \xi^3 / \tilde a^2$,
even in the asymptotic critical region $\xi \gg \tilde a$. This is
the crucial point of our argument.

Thus, in order to have a universal ($\tilde a\;$independent)
scaling form of $C(x)$ for large $|x|\gg\xi$ at fixed $T$
above $T_c$ where $C(x)$ has an exponential form, it is
inescapable to employ $\xi_1$ rather than $\xi$
as the appropriate reference length. Correspondingly, for any $T >
T_c$, there exists an infinitely large region $|x| \gtrsim 24\; \xi^3 /
\tilde a^2$ where the anisotropy of $G(\bf x)$ is no longer a
negligible correction to the isotropic part.
In the critical region $\tilde a \ll |\bf x|\ll \xi$,
on the other hand, where $G(\bf x)$ has a {\it{power-law behavior}}
the nonuniversal part of the difference between $\xi$ and $\xi_{\bf e}$
can be considered as a
negligible non-asymptotic $\it{additive}$ correction. The natural
reference lengths in these two
regions are $\xi$ and $\xi_{\bf e}$, respectively, and a complete
scaling description should embody a kind of crossover in the
scaling variable from $|{\bf x}| / \xi$ to ${\bf x} / \xi _{\bf e}$.
The same situation will arise
in the finite-size problem with respect to the $L$ dependence of
the susceptibility
that is analyzed in the next Section.

The analysis of this Section can be extended to the continuum
version of the $\varphi^4$ theory. In a separate paper \cite
{chen-dohm} we shall show that the results depend on
the cutoff procedure. An (isotropic) exponential large-
$|{\bf x}|$ behavior of $G({\bf x})$ is found for a
smooth cutoff whereas a sharp cutoff implies a nonuniversal
non-exponential $|{\bf x}|$-dependence of $G({\bf x})$. In the
case of a smooth cutoff it is also found that the exponential
correlation length differs from the second-moment correlation
length even though the continuum system is isotropic.

\newpage

\renewcommand{\theequation}{3.\arabic{equation}}
\setcounter{equation}{0}

\section{Lattice effect on the finite-size scaling behavior for $n
\rightarrow \infty$}

We consider the lattice Hamiltonian (\ref{gleichung10}) for a
finite hypercubic geometry with volume $V = L^d$ and with periodic
boundary conditions. We are interested in the exact large-$L$
behavior of the susceptibility $\chi = \hat G (\bf 0)$ above $T_c$
in the large-$n$ limit at finite lattice spacing $\tilde a$.
Specifically we wish to identify the reference length that governs
the expected exponential $L$-dependence at fixed $T > T_c$. The
answer is not clear a priori since in the sum

\begin{equation}
\label{gleichung32} \chi = \frac {\tilde a^d}{n} \sum_j <\varphi_j
\; \varphi_0>
\end{equation}

there are contributions from $<\varphi_j \;\varphi_0>$ in all
directions involving all anisotropic correlation lengths $\xi\bf_
e$ discussed in the preceding Section. The finite-size effect on
$\chi$ at finite lattice constant $\tilde a$ has already been
calculated previously \cite{chen-dohm-b10-1999,
chen-dohm-c9-1073-1998,chen-dohm-b5-1998} where it was expressed in terms of the
second-moment correlation length $\xi$. Here we shall demonstrate
that $\xi_1$ as calculated in the preceding Section, rather than
$\xi$, is the appropriate reference length in the finite-size
scaling structure.

We start from the implicit equation for $n \rightarrow\infty$ at
fixed $u_0n$ \cite{chen-dohm-b5-1998}

\begin{equation}
\label{gleichung33} \chi^{-1} = r_0 + 4u_{0}n L^{-d} \sum_{\bf k}
\left (\hat J_{\bf k} + \chi^{-1}\right) ^{-1}
\end{equation}

which can be rewritten as

\begin{eqnarray}
\label{gleichung34} \chi^{-1} = r_0 - r_{0c} \; + \; 4u_{0}n\; D
\left(\chi^{-1}, L, \tilde {a}\right)\nonumber\\[10pt]
 -\; 4u_{0}n\; \chi^{-1} \int\limits_{\bf k} \; \left [\hat J_{\bf
 k}\left (\hat J_{\bf k} + \chi^{-1}\right)\right]^{-1}
\end{eqnarray}

where $r_{0c} = - 4u_{0}n \int_{\bf k} \; \hat J_{\bf k} ^{-1}$.
The finite-size effect is contained in the function

\begin{equation}
\label{gleichung35} D \left(\chi^{-1}, L, \tilde {a}\right) =
L^{-d} \sum _{\bf k} \left (\hat J_{\bf k} + \chi^{-1}\right)^{-1}
- \int\limits_{\bf k}\left (\hat J_{\bf k} + \chi^{-1}\right)^{-1}
\end{equation}

\begin{equation}
\label{gleichung36} =  \int\limits ^\infty_0 d{\tilde s} \;e\;
^{-{\tilde s}/\chi}\left\{L^{-d} \sum_{\bf k}\; e\; ^{-\tilde s
\hat J_{\bf k}} - \int\limits_{\bf k} \;e\;^{-\tilde s \hat J_{\bf
k}}\right\}\;.
\end{equation}

The summations run over discrete $\bf k$ vectors with components
$k_j = 2\pi m_j/L, m_j = 0, \pm 1, \pm 2, ..., j = 1, 2, ..., d$,
in the range $-\pi/\tilde a \leq k_j < \pi/\tilde a$.
Since $\hat J _{\bf k}$ is a periodic function of each component
$k_j$ the sum in (\ref{gleichung36}) satisfies the Poisson
identity \cite{brezin-1982, morse-1953}

\begin{equation}
\label{gleichung37} L ^{-d} \sum_{\bf k}\; e\; ^{-\tilde s \hat
J_{\bf k}} = \sum_{\bf n} \int\limits_{\bf k}\;e\;^{-\tilde s \hat
J_{\bf k}}\;e\;^{i{\bf {kn}}L}
\end{equation}

where ${\bf{k{\cdot}n}} = \sum_{j}k_{j}n_{j}$. The sum $\sum_{\bf
n}$ runs over all integers $n_j, j=1, 2, ..., d$ in the range $-
\infty \leq n_j \leq \infty$ whereas $\sum\bf_k$ and $\int_{\bf
k}$ have finite cutoffs $\pm \pi/\tilde a$. For the case of a
nearest-neighbor coupling $J > 0$ we have

\begin{equation}
\label{gleichung38} e^{-\tilde s \hat J_{\bf k}} = \prod\limits
^{d} _{m=1}\;\exp \left\{-\frac {4J{\tilde s}} {{\tilde
a}^{2}}\left [1 - \cos (k_m\tilde a)\right]\right\}\;.
\end{equation}

This leads to the representation in terms of the Bessel functions
$I_\nu(z)$, (\ref{gleichung23}),

\begin{equation}
\label{gleichung39} \int\limits_{\bf k}\;e\;^{-\tilde s \hat
J_{\bf k}} = \left [\tilde a^{-1} \; e^{-2s} I_0 (2s)\right]^{d},
\end{equation}

\begin{equation}
\label{gleichung40} L^{-d} \sum_{\bf k}\; e\; ^{-\tilde s \hat
J_{\bf k}} = \left [\tilde a^{-1} \; \sum ^{\infty} _{n=-\infty}
e^{-2s} I_{\nu_n}(2s)\right]^{d}
\end{equation}

where $s = 2\tilde{s}J/\tilde{a}^2$ and where $\nu_n = nL/\tilde a$ are
integers. The resulting expression for $D$ reads

\begin{eqnarray}
\label{gleichung41} D \left(\chi^{-1}, L, \tilde a\right) &=&
\frac {\tilde a^{2-d}}{2J} \int\limits ^\infty_0 ds \;
e\;^{-\tilde a^2 s/(2J\chi)}\; e\; ^{-2ds} \nonumber\\ &\times&
\left\{\left[I_0(2s) + \sum ^{\infty} _{n=1}2 I_{\nu_n}
(2s)\right]^d - \left [I_0(2s)\right]^{d}\right\}\;.
\end{eqnarray}

The large-$L$ limit corresponds to large integers $\nu_n =
nL/\tilde a$. Since we consider this limit at fixed temperature
above $T_c$ we may replace $\chi(t, L)$ in the exponent of (\ref{gleichung41})
by the bulk value $\chi_b = \xi^2/2J$. For the asymptotic behavior
of $I_\nu (2s)$ we refer to Appendix A. The leading term for $L\gg\xi$
comes from the $n=1$ contribution in (\ref{gleichung41}),

\begin{eqnarray}
\label{gleichung42} D \left(\chi^{-1}, L, \tilde a\right) =
\left( d\; \tilde a ^{2-d}/J\right)\int\limits ^\infty_0 ds \;
e\;^{-(\tilde{a}/\xi)^{2} s} \; e\; ^{-2ds}\left[
I_0(2s)\right]^{d-1}I_{L/\tilde a} (2s)\;.
\end{eqnarray}

This integral is identical with that of the bulk correlation
function (\ref{gleichung26}), except that the argument $x/\tilde a$ in
(\ref{gleichung26}) is replaced here by $L/\tilde a$. Therefore the large-$L$
behavior of (\ref{gleichung42}) is analogous to the large-$|x|$ behavior of
(\ref{gleichung27}),

\begin{equation}
\label{gleichung43} D \left(\chi^{-1}, L, \tilde a\right) =
\frac {d\tilde{a}^{2-d}}{2J} (2\pi L/\tilde a)^{(1-d)/2}
\left[\sinh (\tilde a/\xi_1)\right]^{(d-3)/2} \;e\;^{-L/\xi_{1}}
\left[1 + O (L^{-1})\right]
\end{equation}

for $L\gg\xi_1$ where $\xi_1$ is the exponential bulk correlation
length in the direction of one of the cubic axes as determined by
(\ref{gleichung29}). The exact parallelism between the large-distance
behavior of the correlation function $C(x)$ and the large-$L$
behavior of the susceptibility $\chi(t,L)$ is the central result
of this paper. On physical grounds it is quite plausible that for
$L \gg \xi$
there exists a sensitivity of finite-size effects to the length
$\xi_1$ governing the large-distance decay of $C(x)$ rather
than to an averaged length as represented by the second moment
$\xi$ of the correlation function.

The result (\ref{gleichung43}) can be extended to a
$d$-dimensional system with partially finite geometry that is
confined in $\tilde d$ dimensions and infinite in $d - \tilde
d$ dimensions. In this case the prefactor $d$ in
(\ref{gleichung43}) is replaced by $\tilde d$. For $\tilde d = 1$
(film geometry) we find agreement with the result of Barber and
Fisher \cite{barber-1973, agreement}. The authors did not recognize, however,
that their quantity $\Gamma_d (T) = 2\; \rm{arsinh}\left (\Phi_0^{1/2} /
2 \right)$ is identical with the inverse of the exponential
${\it bulk}$ correlation length $\xi_1 / \tilde a$.

The result (\ref{gleichung34}) together with (\ref{gleichung43})
is still valid for arbitrary $\xi_1/\tilde a$ even well above
$T_c$.
Using the known expression for the bulk susceptibility $\chi_b$
\cite{chen-dohm-b10-1999} we obtain from (\ref{gleichung34}) and
(\ref{gleichung43}) the relative deviation from
the bulk critical behavior for $L\gg\xi_{1}\gg\tilde a$ as

\begin{equation}
\label{gleichung44} \Delta\chi \; \equiv \frac {\chi_b-\chi}
{\chi_b}\; = g(L/\xi_1)
\end{equation}

with the universal function in the large-$n$ limit for $2 < d < 4$

\begin{equation}
\label{gleichung45}g(L/\xi_1) = 2\;d\;\pi^{1/2} \left[\Gamma
\left((4-d)/2\right)\right]^{-1} \left(2\xi_1/L\right)^{(d-1)/2}
\;e\;^{-L/\xi_1}\;.
\end{equation}

This result agrees with and goes beyond our previous result in
(132) - (134) or (135) of Ref. \cite{chen-dohm-b10-1999} which
was expressed in terms of $\xi$ rather than $\xi_1$ (compare also
(\ref{gleichung3}) and (\ref{gleichung8}) of the present paper).
Previously we did not yet
recognize the physical origin of the non-scaling contribution $R
(L/\xi, \tilde a/\xi)$ in (134) of Ref. \cite{chen-dohm-b10-1999}.
Now we see that $\xi_{eff}$ as defined in the paragraph after (107)
of Ref. \cite{chen-dohm-b10-1999} turns out to be
identical with $\xi_1$. This is parallel to the bulk
order-parameter correlation function of Section 3.
Thus our previous interpretation in terms
of a violation of finite-size scaling below four dimensions in the
region $L \gg \xi$ was incomplete for the
lattice system (but not for the continuum system with a sharp
cutoff \cite{chen-dohm-b10-1999, chen-dohm, chen-dohm-b7-1999}).
In the critical region $\xi \gg L$, on the other hand, the natural
reference length remains to be $\xi$ and not $\xi_1$. In this
region the function $g(x)$ has a power-law form (not exponential)
in which the non-universal part of the difference between
$L/\xi_1$ and $L/\xi$ can be considered as a negligible correction.

We conclude that nonuniversal and non-negligible lattice effects do
exist in the region $L \gg \xi$ but they can be
absorbed in the finite-size scaling argument by employing
the exponential bulk correlation length.
This remedies the apparent violation of finite-size
scaling found previously below four dimensions
\cite{chen-dohm-b10-1999} and simplifies the physical picture
of critical behavior in confined lattice systems with periodic
boundary conditions. Nevertheless we maintain that a scaling
description of the entire region $0 \leq L/\xi \leq \infty$
requires to embody in the scaling function a kind of crossover in
the finite-size scaling variable from $x = L/\xi$ for $0 \leq
x \lesssim O (1)$ to $x = L/\xi_1$ for $O (1) \leq x \leq
\infty$.

So far our conclusions have only been
shown to be correct in the spherical (large-$n$) limit for $2 < d
< 4$ (and in one-loop order, see Section 4, see also Section 6 for the $d=1$
Ising model). We note that Eq. (\ref{gleichung45}) has a finite
limit also for $d \rightarrow 2$ at fixed $\xi_1$.
Further work is needed to prove whether
finite-size scaling for lattice systems with periodic
boundary conditions below four dimensions is indeed an
asymptotically exact property for finite $n$ beyond one-loop
order. General renormalization-group arguments \cite{zinn-justin-1996,
brezin-1982} are not sufficient for such a proof, as shown in Sect. 3 of Ref.
\cite{chen-dohm-b10-1999}.

For a corresponding analysis of the large-$L$ behavior above $T_c$
within the continuum $\varphi^4$ theory we refer to Refs.
\cite{chen-dohm-b10-1999, chen-dohm, chen-dohm-b7-1999}. In this
case an exponential size dependence of $\Delta\chi$ is found only
for a smooth cutoff whereas a sharp cutoff implies a nonuniversal
non-exponential $L$ dependence of $\Delta\chi$
\cite{chen-dohm-b10-1999, chen-dohm, chen-dohm-b7-1999}.

\newpage

\renewcommand{\theequation}{4.\arabic{equation}}
\setcounter{equation}{0}

\section{Perturbative treatment of finite-size effects for $d<4$}

In this Section we present two different perturbative treatments
of the finite-size effects of the lattice model (\ref{gleichung10})
for finite $n$. We shall focus our interest on $\Delta \chi$ in
the region $L \gg \xi$ above $T_c$ where lattice effects are
expected to be non-negligible according to the exact results
of the preceding Section. In particular we show that only the
first version of the perturbative treatment (in Subsection 4.1)
correctly predicts the exponential size-dependence of $\Delta\chi
\propto e^{-L/\xi_1}$.

\subsection{Ordinary perturbation theory}

First we use ordinary perturbation theory with
respect to $u_0$ without separating the lowest $(\bf {k} =
\bf{0})$ mode of $\varphi(\bf x)$. In one-loop order above $T_c$
the inverse (bare) susceptibility of the lattice model
(\ref{gleichung10}) in a cubic geometry with periodic boundary
conditions is given by \cite{chen-dohm-b10-1999}

\begin{equation}
\label{gleichung50} \chi^{-1} = J_0 \xi^{-2} \left[1 + 4 (n +
2)\;
u_0\; J_0^{-2}\; \xi^{2} \tilde D\; (\xi, L, \tilde a) \; + \; O (u
^{2}_{0})\right]
\end{equation}

with

\begin{equation}
\label{gleichung51} \tilde D = L^{-d} \; \sum_{\bf k}
\left(\xi^{-2} + \hat J_{\bf k} / J_0\right)^{-1}\; -\; \int\limits_{\bf
k}\left (\xi^{-2} + \hat J_{\bf k} / J_0\right)^{-1}
\end{equation}

where $\xi$ is the second-moment correlation length. The function

\begin{equation}
\label{gleichung52} \tilde D (\xi, L, \tilde a) = J_0\; D (J_0\;
\xi^{-2}, L, \tilde a)
\end{equation}

can be represented in terms of Bessel functions according to
(\ref{gleichung41}) and (\ref{gleichung42}). Eqs.
(\ref{gleichung50}) - (\ref{gleichung52}) are valid for general $d
> 2$. Because of the $\bf k
= \bf 0$ term in the sum of (\ref{gleichung51}) the perturbative expression
(\ref{gleichung50}) is not applicable to the region $\xi \gg L$.
In this region a separation of the lowest mode from the higher
modes is necessary (see Sect. 4.2).
But here we are interested in the region $L \gg\xi$
where the function $\tilde D$ is well behaved according to
(\ref{gleichung43}). Applying the RG procedure of Ref.
\cite{chen-dohm-b10-1999} to the bare expression (\ref{gleichung50})
and using the asymptotic form (\ref{gleichung43}) leads to the
scaling result for $d < 4$ and for $L \gg \xi_1 \gg \tilde a$

\begin{equation}
\label{gleichung53} \Delta\chi = g (L/\xi_1) = 4 (n + 2) u^*\;d\;
(2\pi L/\xi_1)^{(1-d)/2}\; e\; ^{-L/\xi_1}\; + O (u^{*2})
\end{equation}

where $u^*$ is the fixed point value of the renormalized coupling
\cite{relation} and where $\xi_1$ is the exponential bulk correlation length given by
(\ref{gleichung29}), up to two-loop corrections. Eq.
(\ref{gleichung53}) has the same form as (\ref{gleichung45}). It
also agrees with and goes beyond our previous perturbative result
(106) of Ref. \cite{chen-dohm-b10-1999}. Now we see that it is
$\xi_1 = \xi/[1 - (\tilde a/\xi)^2 / 24 + ...]$ rather than $\xi$
that should be employed in the scaling representation, similar to
the case $n \rightarrow \infty$ discussed in the preceding
Section. Thus the interpretation of Ref. \cite{chen-dohm-b10-1999}
in terms of a violation of finite-size scaling was incomplete since
the lattice constant $\tilde a$ can be absorbed in $\xi_1$ in a
natural way. The conclusions drawn in the preceding Section after
(\ref{gleichung45}) regarding the validity of finite-size scaling
in terms of $\xi_1$ apply also to finite $n$, at least in one-loop
order. (Beyond one-loop order we expect that the exponential part
of (\ref{gleichung53}) contains the bulk correlation length
$\xi_1$ whose universal amplitude ratio $\lim_{t\rightarrow 0 +} \xi_1 /
\xi$ is slightly larger than 1 \cite{tarko-1975, campostrini-1998}.)
Clearly these conclusions can be extended to a
$d$-dimensional system with partially finite geometry that is
confined in $\tilde d$ dimensions and is infinite in $d-\tilde d$
dimensions. In this case the result (\ref{gleichung53}) remains
valid except that the prefactor $d$ should be replaced by $\tilde
d$ \cite{chen-dohm-b10-1999}. The result of this Subsection will
be extended to the case of the continuum $\varphi^4$ theory in a
separate paper \cite{chen-dohm}.

\subsection{Separation of the lowest mode}

In the following we discuss the result for $\Delta\chi$ for
$L\gg\xi$ if the standard finite-size perturbation theory
\cite{brezin-1985, rudnick-1985, esser-1995, chen-dohm-c9-1073-1998}
is used. The details of the calculation are given in Appendix C. In this
approach the lowest mode is separated and treated exactly whereas
the higher modes are treated perturbatively. Accordingly we decompose

\begin{equation}
\label{gleichung54} \varphi_j = \Phi + L^{-d} \sum_{{\bf k} \neq
{\bf 0}}
\;e\;^{i\;{\bf k\;x}_j}\; \hat \varphi_{\bf k}
\end{equation}

and $H = H_0 + H'$ with

\begin{equation}
\label{gleichung55} H_0 (\Phi) = L^d \left(\frac{1}{2}\; r_0\; \Phi^2 + u_0
\;\Phi^4\right)\;.
\end{equation}

The susceptibility (\ref{gleichung32}) is expressed as

\begin{equation}
\label{gleichung56} \chi = \frac {1} {n}\; \left < \Phi^2 \right >
= \frac {L^d}{n}\; \int\limits d^n \Phi\:\Phi^2 \; P (\Phi)
\end{equation}

where

\begin{equation}
\label{gleichung57} P (\Phi) = \exp \left [- H^{eff} (\Phi)\right ]
 / \int\limits d^n\Phi\;\exp\; \left [- H^{eff} (\Phi)\right ]
\end{equation}

is the order-parameter distribution function with the effective
Hamiltonian

\begin{equation}
\label{gleichung58}H^{eff} (\Phi) = H_0 (\Phi) + \Gamma_0
(\Phi)\;.
\end{equation}

The present approach consists of a perturbative expansion of
$\Gamma_0(\Phi)$ in the $\it exponent$ of $P (\Phi)$ and not
of an expansion of $\chi$ itself. It turns out that this
approach does not capture the correct
(exponential) size dependence of $\Delta\chi$ for $L \gg \xi$ but
instead yields

\begin{equation}
\label{gleichung59}\Delta\chi\propto L^{-d}
\end{equation}

{\it{in any finite order}} of perturbation theory (see Appendix C).

The failure of this approach is due to the
fact that the separation of the zero mode \cite{brezin-1985, rudnick-1985,
esser-1995, chen-dohm-c9-1073-1998} is inadequate in the
region $L \gg \xi$. In this region the zero mode does not have a
dangerous character and all modes including the ${\bf k} = {\bf
0}$ mode should be treated in the same way. This argument is valid
for general $d > 2$ including $d > 4$ (Sect. 5). The amplitude $A
(u_0)$ of the spurious power law $\Delta\chi \propto A(u_0) L^{-d}$
is only partially cancelled $\it order\; by\; order$ in a
perturbative treatment of the higher modes but $A(u_0)$ remains
nonzero at any finite order of perturbation theory.
A complete cancellation of $A(u_0)$ is achieved only in an $\it exact$
treatment of the $\bf {k} \neq \bf {0}$ modes as can be seen from
the exact solution for $n \rightarrow \infty$ (see Eqs.
(12) and (21) of Ref.
\cite{chen-dohm-b7-1999}). These considerations are insensitive to
the lattice spacing and remain valid also within the continuum
$\varphi^4$ theory \cite{chen-dohm}.

We conclude that the perturbative calculation of finite-size
effects above and at $T_c$ requires two different approaches
depending on whether $0 \leq L/\xi \lesssim O (1)$ or $O (1)
\lesssim L/\xi \leq\infty$. In the former case the separation of
the lowest mode is appropriate. In the latter case which includes
the approach to the bulk limit at fixed $T > T_c$ one should
employ the ordinary perturbation approach of Subsection 4.1 where
$\it all$ modes are treated perturbatively. To combine the results
of both approaches requires some matching in an intermediate range
of $L/\xi$. The good agreement of our previous finite-size
calculations \cite{esser-1995} with highly accurate Monte-Carlo
data for the three-dimensional Ising model
\cite{chen-dohm-talapov-1996} was restricted to the region $0 \leq
L/\xi \lesssim O (1)$ whereas the region $O (1) \lesssim L/\xi
\leq\infty$ was not investigated. The exponential size dependence
in the latter region is not correctly included in the results of Ref.
\cite{esser-1995}. The same criticism applies to other finite-size
calculations in the literature which are based on the separation
of the lowest mode. Although these exponential effects are small
they are detectable and clearly distinguishable from power-law
terms as has been demonstrated very recently by Monte-Carlo
simulations for the magnetization of the two- and
three-dimensional Ising model \cite{chen-dohm-stauffer}.

\newpage

\renewcommand{\theequation}{5.\arabic{equation}}
\setcounter{equation}{0}

\section{Finite-size effects for $d > 4$}

In this Section we extend our study of finite size effects to $d >
4$ within the lattice model (\ref{gleichung10}) for cubic geometry
and periodic boundary conditions. We focus our interest on the
approach of the susceptibility $\chi$ to the bulk susceptibility
$\chi_b$ above $T_c$. To provide a correct description of this
approach is a basic task of finite-size theory. This
corresponds to the region $L \gg \xi$
where the exponential correlation length $\xi_1$ is expected to
become an important length scale according to the results of the
preceding Sections.
We shall show that, in addition to $\xi_1$, the second reference
length \cite{chen-dohm-c9-1073-1998, chen-dohm-b5-1998,
chen-dohm-a251-1998} $l_0 \sim u_0^{1/(d-4)}$ associated with the
higher (inhomogeneous) modes
remains relevant for the large-$L$ behavior of $\chi$ and that a
single-variable (lowest-mode) finite-size scaling description \cite{brezin-1985,
binder-1985} of $\chi$ fails for $L \gg \xi$.

\subsection{Exact results for $n \rightarrow\infty$}

For $\chi_b \gg \tilde a^2$ the inverse bulk susceptibility
above $T_c$ for $4 < d < 6$ and $n \rightarrow \infty$ at fixed
$u_0n$ is determined by

\begin{equation}
\label{gleichung5.2} \chi_b^{-1} = r_0 - r_{0c} - 4u_0n
\;\chi_b^{-1} \int\limits_{\bf k} \hat J_{\bf k}^{-2}\left\{1 +
O \left[(d-4)^{-1} \left (\chi_b^{-1} \tilde a^2
\right)^{(d-4)/2}\right]\right\}
\end{equation}

as follows from (\ref{gleichung34}) for $L \rightarrow \infty$. From
(\ref{gleichung5.2}) and (\ref{gleichung34}) we then obtain the
leading relative deviation of $\chi$ from $\chi_b$ for
$\Delta\chi\ll1$ as

\begin{equation}
\label{gleichung5.3} \Delta\chi\equiv \frac {\chi_b-\chi} {\chi_b}
= l_0^{d-4}\; \chi_b \;J_0^2 D(\chi_b^{-1}, L, \tilde a) + O
\left[(\Delta\chi)^2\right]
\end{equation}

with the reference length \cite{chen-dohm-b5-1998}

\begin{equation}
\label{gleichung5.4}l_0 = \left[\frac {4u_0n}{J_0^2 \left(1 +
4u_0n \int\limits_{\bf k} \hat J_{\bf k}^{-2}\right)}
\right]^{1/(d-4)}\;.
\end{equation}

The bulk susceptibility can be expressed in terms of the
second-moment correlation length $\xi$ as $\chi_b = J_0^{-1}
\;\xi^2$. Using the large-$L$ behavior of $D (\chi_b^{-1}, L,
\tilde a)$ according to (\ref{gleichung43}) we obtain from
(\ref{gleichung5.3}) for $L\gg\xi$

\begin{equation}
\label{gleichung5.5} \Delta\chi\sim d (2 \pi)^{(1-d)/2}
(L/l_0)^{4-d} (L/\xi)^{-2} \left[(L/\tilde a) \sinh (\tilde
a/\xi_1)\right]^{(d-3)/2} e^{-L/\xi_1}
\end{equation}

where now the exponential bulk correlation length $\xi_1$,
(\ref{gleichung29}), governs the exponential size dependence,
similar to the case $d < 4$. In the asymptotic region $\xi\gg
\tilde a$ we may replace $\xi$ by $\xi_1$ in the non-exponential
part of (\ref{gleichung5.5}). This yields the two-variable
finite-size scaling form

\begin{equation}
\label{gleichung5.6} \Delta\chi = g \left(L/\xi_1,
(L/l_0)^{4-d}\right)
\end{equation}

with the exact scaling function for $L \gg \xi \gg \tilde a$

\begin{equation}
\label{gleichung5.7} g (x, y) =\;d\; (2 \pi)^{(1-d)/2}\; y\;
x^{(d-7)/2}\; e^{-x}\;.
\end{equation}

Unlike the corresponding scaling function $g(x)$ in
(\ref{gleichung45}) for $d<4$ we see that here we need $\it{two}$
scaling variables $x = L/\xi_1$ and $y = (L/l_0)^{4-d}$. In the
present context where the lowest mode plays no particular role, the
second variable $y$ is associated with the higher modes and has
nothing to do with the dangerous character of $u_0$. The present result
(\ref{gleichung5.6}) and (\ref{gleichung5.7}) complements our
previous two-variable finite-size scaling function (138) - (142)
for the lattice model in Ref. \cite{chen-dohm-b5-1998} where $\xi$
instead of $\xi_1$ was employed \cite{replace}. A complete description of
the scaling form of $\chi$ in the entire (asymptotic) $L^{-1} - \xi^{-1}$ plane
requires to incorporate in $g(x, y)$ a kind of crossover from the
variables $(L/\xi, y)$ for the region $0 \leq L/\xi \lesssim O(1)$
to $(L/\xi_1, y)$ for the region $O (1) \lesssim L/\xi_1
\leq\infty$.

We recall that an alternative choice of the scaling variables
$(L/\xi, y)$ is $(w, y)$ where \cite{chen-dohm-c9-1073-1998}

\begin{equation}
\label{gleichung5.8} w = (L/\xi)^2 y^{-1/2} = t (L/\tilde\ell)^{d/2},\;
\;\tilde\ell = l_0 (\xi_0/l_0)^{4/d}\;.
\end{equation}

Correspondingly the susceptibility can be represented as
\cite{chen-dohm-c9-1073-1998}

\begin{equation}
\label{gleichung5.8a} \chi = L^{d/2}\; \tilde P(w, y)\;.
\end{equation}

Instead of $(w, y)$ an equivalent choice is $(w^{2/d}, y)$ where

\begin{equation}
\label{gleichung5.9} w^{2/d} = L/\ell_T
\end{equation}

contains Binder's "thermodynamic length" $\ell_T$ \cite{binder2}
which is related to $\xi_0$ and $l_0$ as

\begin{equation}
\label{gleichung5.10} \ell_T = l_0^{(d-4)/d}\; \xi^{4/d} = l_0^{(d-4)/d} \; \xi_0^{4/d} \;
t^{-2/d}\;.
\end{equation}

This length scale, together with $l_0$, plays an important role in
the region $0 \leq L/\ell_T \lesssim O(1)$ where the dangerous
character of $u_0$ is important (corresponding to
the region between the curved lines in Fig. 1 of Ref.
\cite{chen-dohm-c9-1073-1998}) but $\ell_T$ loses its significance
outside this region. In particular in the region $O(1) \lesssim
L/\xi \leq\infty$ the correlation length $\xi_1$ (and $\xi$) and
the reference length $l_0$ associated with the higher modes
\cite{chen-dohm-c9-1073-1998}
govern the finite-size effects, as
demonstrated by (\ref{gleichung5.5}) - (\ref{gleichung5.7}). Thus,
not $\ell_T$ alone but $\xi_1$ and $l_0$ are indispensable
for a complete description of the finite-size effects in the
entire asymptotic (large $L$, small $t$) region for $d>4$.
In the following Subsection we show that ignoring the lengths $\xi$
or $\xi_1$ and $l_0$
implies an incorrect large-$L$ dependence of $\chi$ at
any fixed $T>T_c$.

\subsection{Lowest-mode approximation for $n \rightarrow \infty$}

Neglecting the $\bf k \neq \bf 0$ contributions to $\chi$ leads to

\begin{equation}
\label{gleichung5.11} \chi_0 = 2 \left[r_0 + \left(r_0^2 + 16 u_0n
\;L^{-d}\right)^{1/2}\right]^{-1}
\end{equation}

for $n \rightarrow\infty$ at fixed $u_0n$. In this approximation
we have $\chi_b^0 = r_0^{-1} = (a_0 t)^{-1}$.

This yields

\begin{eqnarray}
\label{gleichung5.12} \Delta\chi_0 \equiv \frac {\chi_b^0 -
\chi_0}{\chi_b^0}
& = & 1 - 2 \left[1 + \left(1 + 16
u_0n\; r_0^{-2} \; L^{-d}\right)^{1/2}\right]^{-1}
\end{eqnarray}

In the present approximation the lengths $\xi_0$ and $l_0$
are reduced to

\begin{equation}
\label{gleichung5.14} \xi_0 = (J_0/a_0)^{1/2}\;,\; l_0 = \left(4 u_0n \;
J_0^{-2}\right)^{1/(d-4)}\;.
\end{equation}

Thus $\Delta\chi_0$ can be expressed in terms of the thermodynamic
length $\ell_T$, (\ref{gleichung5.10}), as

\begin{eqnarray}
\label{gleichung5.15} \Delta\chi_0 & = & 1 - 2 \left\{1 + \left[1
+ 4 \left(\ell_T/L\right)^{-d}\right]^{1/2}\right\}^{-1}
\end{eqnarray}

\begin{eqnarray}
\label{gleichung5.15a}& = & \left(L/\ell_T\right)^{-d} + O
\left(L^{-2d}\right)
\end{eqnarray}

for $L \gg \ell_T$. Comparison of (\ref{gleichung5.15})
with (\ref{gleichung5.5}) -
(\ref{gleichung5.7}) shows that the lowest-mode approximation
\cite{brezin-1985} fails both with regard to the $L$ dependence of
$\Delta\chi$ as well as with regard to the temperature dependence
of the reference length scale $\ell_T \neq \xi_1$ in the scaling
variable.

The same criticism applies to the phenomenological
single-variable scaling form $\Delta\chi = f (L/\ell_T)$ proposed by
Binder et al. \cite{binder-1985}. The recent statement \cite{binder-preprint}
that the single-variable scaling form is presumably
true asymptotically for $L \rightarrow \infty$ is correct for $T =
T_c$ (more precisely, for $L \rightarrow \infty$ at fixed finite
$w$). Furthermore, the lowest-mode result $\chi_0 (t, L) = L^{d/2}
\tilde P (w, 0)$ correctly contains the limit $\chi_b^0 (t) =
\chi_0 (t, \infty)$ of $\chi_0(t, L)$ for $L \rightarrow\infty$
at fixed $L/\xi$ (see Eq. (104) of \cite{chen-dohm-c9-1073-1998}
and Eq. (102) of \cite{chen-dohm-b5-1998}) but does not correctly
describe the size dependence in {\it approaching} this limit
$\chi_b^0(t)$ (see Eq. (104) of \cite{chen-dohm-b5-1998}).
A corresponding statement is also true with regard to
the magnetization below $T_c$ as confirmed by Monte Carlo
simulations for the $d = 5$ Ising model \cite{cheon}.
As pointed out in \cite{cheon}, the property

\begin{equation}
\label{gleichung15.b}\lim_{L\to\infty} \chi (t, L) / \chi_b (t)\; =\; 1
\end{equation}

{\it{at arbitrary fixed}} $L/\xi$ is a nontrivial feature that is valid
only for $d > 4$ and that is correctly contained in the
lowest-mode approximation and in the Binder et al. scaling form
whereas for $d < 4$ the same limit yields the {\it{function}} $f
(L/\xi) \neq 1$ for $L/\xi < \infty$.

For any fixed $T \neq T_c$, however, the leading size dependence $\propto
L^{-d}$ predicted by the single-variable scaling form
\cite{binder-1985}
is incorrect. The origin for this defect are the missing higher
modes. At fixed $w$ for large $L$, these modes only cause (slowly decaying)
corrections $\sim O(y^{1/2})$ to the leading size dependence
$L^{d/2} \tilde P(w, 0)$ of
the lowest-mode approximation \cite{chen-dohm-c9-1073-1998}.
For fixed $T \neq T_c$ and large $L$ corresponding to $w \gg 1$ and
$y \ll 1$, however, the higher modes and the lowest mode must be treated in the same way,
as shown in Sect. 4.2, and the effects of the higher modes become
increasingly dominant with increasing $L/\xi$ and can no longer be
considered only as corrections.

More specifically, the structure of the scaling function
(\ref{gleichung5.8a}) can be written as

\begin{equation}
\label{gleichung5.16a} L^{d/2}\; \tilde P (w, y) = L^{d/2}\; \tilde P
(w, 0) + \Delta (t, L)
\end{equation}

where $\Delta$ describes the size effect of the higher modes. The
size effect $\Delta_0$ of the zero mode is contained in

\begin{equation}
\label{gleichung5.17a} L^{d/2}\; \tilde P (w, 0) = \chi_b (t) +
\Delta_0 (t, L)
\end{equation}

where $\chi_b(t)$ is the bulk susceptibility. The
crucial point now is that for sufficiently large
$L \gg \xi$ the structure of
$\Delta$ becomes \cite{chen-dohm-b7-1999}

\begin{equation}
\label{gleichung5.18a} \Delta = - \Delta_0 + O (e^{-L/\xi_1})\;,
\end{equation}

i.e., the zero-mode size dependence of $\Delta_0 \propto L^{-d}$ is
exactly cancelled by the higher-mode size dependence of $\Delta$.
Thus, $\Delta$ is not small compared to
$\Delta_0$ and it is inadequate to refer to the size
effects of $\Delta$ only as "corrections to the lowest-mode result"
\cite{binder-preprint, luijten-1999}.

\subsection{Perturbative treatment for finite $n$}

For finite $n$ a perturbative treatment of the finite-size effects
becomes necessary. Our arguments (in Sect. 4) for the necessity of
$\it two$ different perturbative approaches remain valid also for $d >
4$. A one-loop perturbation calculation on the basis of a
separation of the lowest mode was presented recently
\cite{chen-dohm-c9-1073-1998} for the case
$n = 1$. The results of this calculation are applicable to the
region $0 \lesssim L/\xi \lesssim O(1)$ but the quality of this
approach deteriorates with increasing $L/\xi$ in the region $O(1)
\lesssim L/\xi \leq \infty$. For the latter region the following
ordinary perturbation calculation with respect to $u_0$ is appropriate.

The bare perturbative one-loop expression (\ref{gleichung50})
remains valid also for $d > 4$. This leads to

\begin{equation}
\label{gleichung5.16} \Delta\chi = 4(n+2) u_0\;J_0^{-2}\; \xi^2
\;\tilde D (\xi, L, \tilde a) + O(u_0^2)
\end{equation}

where $\tilde D$ is given by (\ref{gleichung52}) and
(\ref{gleichung41}).
Here we interpret the prefactor $4(n+2) u_0\;J_0^{-2}$
as $l_0^{d-4} + O(u_0^2)$ as indicated by the result
(\ref{gleichung5.3}) for $\chi$ in the large-$n$ limit. Using
(\ref{gleichung52}) and (\ref{gleichung43}) for $L \gg\xi$ we
arrive at the same expression for $\Delta\chi$ as given already in
(\ref{gleichung5.5}) where now the reference length $l_0$ for
finite $n$ is

\begin{equation}
\label{gleichung5.1} l_0 = \left\{\frac {4u_0 (n +
2)} {J_0^2 \left[1 + 4u_0 (n + 8) \int\limits_{\bf k}
\hat J_{\bf{k}} ^{-2}\right]} \right\} ^{1/(d-4)}\;.
\end{equation}

The coefficient $4(n+8)$ in the denominator is inferred from the
form (\ref{gleichung5.4}) in the large-$n$ limit and from the
previous result in (30) of Ref. \cite{chen-dohm-c9-1073-1998} for
$n = 1$. (Our present definition of $l_0$ differs from that of
Ref. \cite{chen-dohm-c9-1073-1998} by the factor $[4 (n+2)]^{1/(d-4)}$).
In the asymptotic region $L \gg\xi\gg\tilde a$ we arrive at a
one-loop scaling form of $\Delta\chi$ for finite $n$ which is identical with
(\ref{gleichung5.6}) and (\ref{gleichung5.7}). The $n$ dependence enters only the
expressions for $l_0$ and

\begin{equation}
\label{gleichung5.16b}\xi_0 = (J_0/a_0)^{1/2} \left[1 + 4
(n+2) u_0 \int\limits_{\bf k} \hat J_{\bf k}^{-2}\right]^{1/2}\;.
\end{equation}

For $n=1$ this result for $\Delta\chi$ complements our previous
two-variable finite-size scaling function (97) and (99) in Ref.
\cite{chen-dohm-c9-1073-1998} which did
not yet incorporate the exponential size dependence $\propto
e^{-L/\xi_1}$ for $L\gg\xi$.

Similar to the case $n \rightarrow\infty$, the lowest-mode
approximation \cite{brezin-1985} and the phenomenological
single-variable scaling form \cite{binder-1985} lead to the
incorrect power law for finite $n$

\begin{equation}
\label{gleichung5.17} \Delta\chi_0 = \frac {4(n+2)u_0} {r_0^2}
L^{-d} + O(L^{-2d}) = (L/\ell_T)^{-d} + O (L^{-2d})
\end{equation}

with $\ell_T$ given by (\ref{gleichung5.10}). Our comments after
(\ref{gleichung5.15}) apply also to (\ref{gleichung5.17}). In
summary, while the two-variable scaling form correctly embodies
the $L$-dependent approach $\propto e^{-L/\xi_1}$ to the bulk limit
at fixed $T > T_c$ this crucial information is lost in the
single-variable scaling form and in the lowest-mode approximation.
Therefore the reduction of the two scaling variables to a single
scaling variable is not justified.

Very recently our prediction of the exponential
size dependence (\ref{gleichung5.7}) has been confirmed
by Monte Carlo simulations for the five-dimensional
Ising model \cite{stauffer,chen-dohm-stauffer}.

An analysis of finite-size effects for $d > 4$ within the $\it
{continuum}$ $\varphi^4$ theory will be given in a separate paper
\cite{chen-dohm} where it is shown that the finite-size effects
depend significantly on the cutoff procedure. For a $\it{smooth}$
cutoff the results for the Binder cumulant at $T_c$ and for the
two-variable finite-size scaling function of $\chi$ are found to
be different from those found previously
\cite{chen-dohm-c9-1073-1998, chen-dohm-b5-1998,
chen-dohm-b7-1999, chen-dohm-a251-1998, chen-dohm-c9-1007-1998}
for a $\it sharp$ cutoff, see also the note added in
Ref. \cite{chen-dohm-b10-1999}.

\newpage

\renewcommand{\theequation}{6.\arabic{equation}}
\setcounter{equation}{0}

\section{One-dimensional Ising model}

In this Section we illustrate in an elementary way the close
connection between the exponential bulk correlation length
$\xi_1$ and the finite-size scaling
structure for the example of the exactly solvable one-dimensional
Ising model.
Although the critical temperature $T_c = 0$ vanishes, this model
has well defined correlation lengths $\xi$ and $\xi_1 \neq \xi$
for $T>0$, which diverge for $T \rightarrow T_c = 0$.

First we consider spins $s_j = \pm 1$ on sites $x_j$ of an
$\it{infinite}$ chain with a lattice spacing $\tilde a$. The Hamiltonian
reads

\begin{equation}
\label{gleichung6.1} H = - J \sum ^{\infty} _{j=-\infty} s_j
s_{j_{+1}}\;.
\end{equation}

The correlation function is well known, see e.g. Ref. \cite{baxter-1982}.
The exact result has an exponential form for arbitrary distances
$|x_i - x_j|$,

\begin{equation}
\label{gleichung6.2} <s_i s_j> = \exp (-|x_i - x_j| / \xi_1)
\end{equation}

with the exponential correlation length

\begin{equation}
\label{gleichung6.3} \xi_1 = \tilde a\; [\ln (\lambda_+ /
\lambda_-)]^{-1}
\end{equation}

where $\lambda_+$ and $\lambda_-$ are the eigenvalues of the
transfer matrix \cite{baxter-1982} with $\lambda_+ > \lambda_-$. Obviously
(\ref{gleichung6.2}) has a scaling form in terms of $\xi_1$.

In order to calculate the correlation length $\xi$ as defined in
(\ref{gleichung1}) we consider the Fourier transform

\begin{equation}
\label{gleichung6.4} \hat G (k) = \sum^{\infty} _{j=-\infty} <s_0 s_j> \exp
(-\;i\;k\;j\;\tilde a\;)
\end{equation}

where $s_0$ denotes the spin on a fixed site $x_0$. Using (\ref{gleichung6.2}) and
(\ref{gleichung6.3}) we obtain

\begin{equation}
\label{gleichung6.5} \hat G (k) =
\frac{1-(\lambda_{-}/\lambda_{+})^2}{1+(\lambda_{-}/\lambda_{+})^2
- 2 (\lambda_{-}/\lambda_{+}) \cos k\;\tilde a}\;.
\end{equation}

This leads to the exact result $\hat G (0) = \exp (2\beta\varepsilon)$
and

\begin{equation}
\label{gleichung6.6} \xi^2 = \hat G (0)\frac{\partial}{\partial
k^2}\left[\hat G (k)\right]^{-1}_{|{_{k=0}}} = \frac{\tilde
a^2}{4\left[\sinh(\tilde a/2\xi_1)\right]^2}
\end{equation}

where $\xi_1$ is given by (\ref{gleichung6.3}). This relation between $\xi$
and $\xi_1$ is identical with (\ref{gleichung28}) or (\ref{gleichung29}) which was
derived for the $\varphi^4$ model in the large-$n$ limit in Sect. 3 and in
one-loop order in Sect. 4. In particular we again have $\xi_1 /
\xi\rightarrow 1$ for $T \rightarrow T_c = 0$.

Now we consider the finite-size effect on the susceptibility
$\chi_L$ of a $\it{finite}$ one-dimensional Ising chain which
consists of $N$ spins and which has a length $L = N \tilde a$. We
assume periodic boundary conditions. The partition function is \cite{baxter-1982}
$Z_N = \lambda ^{N}_{+} + \lambda ^{N} _{-}$ which is valid also
at finite magnetic field $h$. The second derivative with respect to
$h$ leads to the exact finite-size scaling form of the relative deviation
from the bulk susceptibility at $h = 0$

\begin{equation}
\label{gleichung6.7} \Delta\chi = \frac{\chi_\infty -
\chi_L}{\chi_\infty} = \frac{2\; e^{-L/\xi_1}}{1 +
e^{-L/\xi_1}}\;.
\end{equation}

Eq. (\ref{gleichung6.7}) is valid for arbitrary $L/\xi_1$ where $\xi_1$ is identical with the
exponential bulk correlation length (\ref{gleichung6.3}).

The crucial point is that $L/\xi_1$ and not $L/\xi$ is the natural
finite-size scaling variable. If $\Delta\chi$ were expressed in terms of $\xi$
then an apparent violation of finite-size scaling would arise from
the $\tilde a$ dependent difference between $\xi_1$ and $\xi$,

\begin{equation}
\label{gleichung6.8} \Delta\chi = 2\; e^{-L/\xi} \exp \left(L \tilde
a^2 / 24 \xi^3 \right)
\end{equation}

for $L \gg \xi \gg \tilde a$,  in the same way as found previously
\cite{chen-dohm-b10-1999} for the
$\varphi^4$ model. Similarly, the bulk scaling form
for the correlation function would be violated if the result
(\ref{gleichung6.2}) would be expressed in terms of
$\xi$. Thus the exact results (\ref{gleichung6.2}),
(\ref{gleichung6.6}) and (\ref{gleichung6.7})
demonstrate in an elementary way the significant difference
between $\xi$ and $\xi_1$ as well as the close connection
between bulk and finite-size scaling.

It would be interesting to extend this analysis to the
exact results for the two-dimensional Ising model
\cite{ferdinand-1969, au-yang-1980} and to compare
these results with the exponential size dependence
found in recent Monte Carlo data in Fig. 2c of
Ref. \cite{chen-dohm-stauffer}.

$\bf {Note\; added}$

The distinction between $\xi$ and $\xi_1$ is significant also for
resolving discrepancies in the interpretation of Monte Carlo
simulation results of percolation phenomena \cite{malarz}.

\subsection*{Acknowledgment}

Support by Sonderforschungsbereich 341 der Deutschen
Forschungsgemeinschaft and by NASA under contract number 1201186 is
acknowledged. One of us (X.S.C.) thanks the National Natural
Science Foundation of China for support under Grant No. 19704005.
We also thank the referees for useful comments.

\newpage

\section*{Appendix A : Anisotropy of the exponential correlation
length}

\renewcommand{\theequation}{A.\arabic{equation}}
\setcounter{equation}{0}

We start from (\ref{gleichung25}) for the case where ${\bf{x}} = (x_1, x_2, 0, ...)$,

\begin{equation}
\label{gleichunga.1} G ({\bf {x}}) = \frac {{\tilde a} ^{2-d}} {2J}
\int\limits ^\infty_0 ds \; e ^{-(\tilde a / \xi) ^{2}s}\; e
^{-2ds} [I_0 (2s)]^{d-2} I_{\nu^{}_{1}} (2s) I_{\nu^{}_{2}}(2s)
\end{equation}

with $\nu_i = |x_i| / \tilde a$. For large $\nu$ and large
$s=\nu z/2$ we have \cite{see-1}

\begin{equation}
\label{gleichunga.2} I_\nu (\nu z) \sim (2 \pi \nu)^{-1/2}
q^{-1/2} \exp \left(\nu \left\{{q + \ln} \left[{z
(1+q)}^{-1}\right]\right\}\right)
\end{equation}

with $q = (1 + z^2) ^{1/2}$. Furthermore we use the large-$s$
behavior \cite{see-2}

\begin{equation}
\label{gleichunga.10} I_0 (2s) = e^{2s} (4\pi s)^{-1/2} \left [1 +
O \left(s^{-1}\right)\right]\;.
\end{equation}

For sufficiently large $\nu_i$ a saddle-point approach suffices to
perform the integration over $s$ and to
determine the exponential large-$|x_i|$ behavior of
(\ref{gleichunga.1}). Introducing the angle $\theta$ according to
$\nu^{}_1 = r \cos \theta, \nu^{}_2 = r \sin \theta$ we obtain the
exponential part of $G(\bf x)$ as

\begin{equation}
\label{gleichunga.3} G (\bf x)\; \sim \; \exp \left\{-|{\bf x}|/
\xi_{\rm 1} (\theta)\right\}
\end{equation}

where $\xi_1 (\theta)$ denotes the anisotropic exponential
correlation length. The angular dependence is given by

\begin{eqnarray}
\label{gleichunga.4} \frac {\tilde a} {\xi_1(\theta)} &=& (\cos
\theta)\ln \left[ u^{1/2} \cos \theta + (1+u \cos^{2}
\theta)^{1/2}\right] \nonumber \\ &+& (\sin \theta) \ln \left [u^{1/2} \sin \theta + (1+u \sin^{2}
\theta)^{1/2}\right]
\end{eqnarray}

where

\begin{equation}
\label{gleichunga.5} u = b (b^2 - 4) \left[b + \left(b^2 \sin^2 2
\theta+ 4 \cos^2 2 \theta \right)^{1/2}\right]^{-1}\;,
\end{equation}

\begin{equation}
\label{gleichunga.6} b = 2 + (\tilde a/\xi)^{2} / 2\;.
\end{equation}

For the case $\theta = 0$ this yields

\begin{equation}
\label{gleichunga.7} \xi_1(0) = \frac{\tilde a}{2} \left
[{\rm{arcsinh}}
\left (\frac {\tilde a}{2\xi}\right ) \right]^{-1}
\end{equation}

corresponding to (\ref{gleichung29}). For the case $\theta =
\pi/4$ the result is

\begin{equation}
\label{gleichunga.8} \xi_1 (\pi/4) = \frac {\tilde a}{2^{3/2}}
\left [{\rm{arcsinh}} \left(\frac {\tilde
a}{2^{3/2}\xi}\right)\right]^{-1}\;.
\end{equation}

Eqs. (\ref{gleichunga.7}) and (\ref{gleichunga.8}) agree
with (4.12) and (4.14) of \cite{fisher-1967} for $d = 2$.
For $\xi \gg \tilde a$ we obtain from (\ref{gleichunga.4}) -
(\ref{gleichunga.6})

\begin{equation}
\label{gleichunga.9} \frac {\tilde a}{\xi_1 (\theta)} = \frac
{\tilde a}{\xi} \left [1 - \frac {1}{48} \left(1 + \cos^2 2 \theta
\right) \left(\frac {\tilde a}{\xi}\right)^2 + O \left(\tilde
a ^4 / \xi^4\right)\right]\;.
\end{equation}

For $\theta = 0$ and $\theta = \pi/4$, (\ref{gleichunga.9})
disagrees with (4.13) and (4.15) of Ref. \cite{fisher-1967}.
Asymptotically $(\xi \rightarrow\infty)$ we obtain from
(\ref{gleichunga.9}) $\xi_1 / \xi \rightarrow 1$ for all $\theta$
(in the spherical limit and in one-loop order). Nevertheless,
because of the exponential form of $G (\bf x)$, the nonasymptotic
$\theta$-dependence remains non-negligible in $G ({\bf x})$ for
$|{\bf x}| \gtrsim 24 \xi^3/\tilde a^2$ even arbitrarily close to
$T_c$, see the reasoning in the context of (\ref{gleichung31}) and
(\ref{gleichung31a}).

To derive the large-$|x|$ behavior (\ref{gleichung27}) for the
case ${\bf x} = (x, 0, ...)$ we use (\ref{gleichunga.2}) and
(\ref{gleichunga.10}). Expanding around the maximum of the
exponential part of the integrand of $C(x)$, (\ref{gleichung26}),
and performing the integration over $s$ leads to (\ref{gleichung27}).

\newpage
\section*{Appendix B : Bulk correlation function in one-loop order for
$d < 4$}

\renewcommand{\theequation}{B.\arabic{equation}}
\setcounter{equation}{0}

In terms of the second-moment correlation length
(\ref{gleichung1}) the bare bulk two-point vertex function at
finite $\bf k$ above $T_c$ is given in one-loop order for $d>2$ by
\cite{chen-dohm-b10-1999}

\begin{equation}
\label{gleichungb.1} \Gamma^{(2)} ({\bf k}, \xi, u_0, \tilde a, d)
= \hat J_{\bf k} + J_0 \xi^{-2} + O (u_0^2),
\end{equation}

corresponding to the integrand of $G(\bf x)$ in the form of
(\ref{gleichung22}) and (\ref{gleichung19}).
Employing the renormalization procedure at finite lattice constant
$\tilde a$ as described in Section 2.2 of Ref.
\cite{chen-dohm-b10-1999} we obtain for $d \leq 4$

\begin{equation}
\label{gleichungb.2} \Gamma^{(2)} ({\bf k}) =
Z^{-1}_{\varphi}\left [\hat J_{\bf k} + J_0 \xi^{-2} + O
\left(u(\ell)^2\right)\right ] \exp \int ^{\ell}\limits_{1}
\zeta_\varphi(\ell^{\prime}) \frac {d \ell^{\prime}} {\ell^{\prime}}\;.
\end{equation}

For the application to $T > T_c$ we choose the flow parameter as
$\ell = \xi_0/\xi$ where $\xi_0$ is the asymptotic amplitude of $\xi$
above $T_c$. For the case ${\bf x} = (x, 0, 0 ...)$ and in the
limit $|x|\gg \tilde a$ the correlation function is in one-loop
order for $d\leq 4$

\begin{equation}
\label{gleichungb.3} G ({\bf x}) = \int \limits_{\bf k} \left [\Gamma^{(2)}
({\bf k})\right] ^{-1}\; e^{i\bf {kx}} = Z_\varphi\; C (x, \xi_1, \tilde a)\exp
\int _{\ell} \limits^{1} \zeta_\varphi (\ell^{\prime}) \frac {d \ell^{\prime}} {\ell^{\prime}}
\end{equation}

with $C (x, \xi_1, \tilde a)$ given by
(\ref{gleichung27})-(\ref{gleichung29}), apart from corrections of
$O \left(u(\ell)^2\right)$, where $u (\ell)$ is the effective
four-point coupling \cite{chen-dohm-b10-1999}. In the asymptotic region
$\xi_1 \gg \tilde a$ we obtain from (\ref{gleichungb.3})

\begin{equation}
\label{gleichungb.4} G ({\bf x}) = Z_\varphi \left[A^{(2)}
\right]^{-1} \xi ^{-\eta}\; \xi_0 ^\eta \;\;C(x, \xi_1, \tilde a)
\left[1 + O \left(u^{*2}\right)\right]
\end{equation}

with

\begin{equation}
\label{gleichungb.5} A^{(2)} = \exp
\left\{\int_1^0\left[\zeta_\varphi(\ell^{\prime}) -\zeta_\varphi(0)\right]
\frac {d \ell^{\prime}} {\ell^{\prime}}\right\}
\end{equation}

and the critical exponent $\eta = - \zeta_\varphi (0)$. In (\ref{gleichungb.4}) the
asymptotic $\left(|x|\gg \xi_1\right)$ form of $C\left(x, \xi_1,
\tilde a\right)$ is

\begin{equation}
\label{gleichungb.6} C \left(x, \xi_1, \tilde a \right)\; = \;  \frac {{\tilde a} ^{2-d}}
{4J} \left (\frac {\tilde a} {2\pi|x|}\right)^{(d-1)/2} \left (\frac {\tilde
a}{\xi_1}\right) ^{(d-3)/2} \; e \; ^{-|x|/\xi_1}\;.
\end{equation}

Asymptotically we may replace $\xi^\eta$ in (\ref{gleichungb.4})
by $\xi_1^\eta \left [1 + O \left(\xi^{-2}\right)\right]$.
Therefore $G (\bf x)$ can be rewritten in the asymptotic scaling
form for $d < 4$

\begin{equation}
\label{gleichungb.7} G ({\bf x}) = \left(\tilde a /|x|
\right)^{d-2+\eta}\;\Phi \left(|x| / \xi_1 \right)
\end{equation}

with the scaling function for $|x|/\xi_1 \gg 1$

\begin{equation}
\label{gleichungb.8} \Phi \left(|x|/\xi_1\right)\; =\; \tilde A\; \frac{{\tilde a}^{2-d}}{{4J(2\pi)}^{(d-1)/2}}
\left(\frac{|x|}{\xi_1} \right)^{\frac{1}{2}(d-3)+\eta} \exp
\left(-|x|/\xi_1\right),
\end{equation}

apart from $O \left (u^{*2}\right)$ corrections. The
amplitude $\tilde A$ is for $d<4$ \cite{chen-dohm-b10-1999}

\begin{equation}
\label{gleichungb.9} \tilde A = Z_\varphi \left(u, \tilde a/\xi_0,
d\right)\; \left(\xi_0/\tilde a\right)^\eta \left[A^{(2)}\right]^{-1}\;.
\end{equation}

\newpage

\section*{Appendix C : Separation of the lowest mode}

\renewcommand{\theequation}{C.\arabic{equation}}
\setcounter{equation}{0}

In the following we show that the separation of the lowest mode
implies $\Delta\chi\propto L^{-d}$ for $L\gg\xi$ and for general
$d > 2$ in any finite order of perturbation theory.
In leading order of the ${\bf k} \neq {\bf 0}$ modes the
effective Hamiltonian (\ref{gleichung58}) becomes
\cite{chen-dohm-c9-1073-1998}

\begin{equation}
\label{gleichung059} H^{eff} (\Phi) = L^d \left [\frac{1}{2}\;
r_0^{eff} \;\Phi^2 + u_0^{eff}\; \Phi^4\right ]\;,
\end{equation}

\begin{equation}
\label{gleichung60} r_0^{eff} = r_0 - r_{0c} + 4(n+2) u_0 \left
[L^{-d} \sum _{{\bf k}\neq {\bf 0}} \left(r_0 - r_{0c} + \hat J_{\bf
k}\right)^{-1} - \int\limits_{\bf k} \hat J_{\bf k}^{-1}\right]\;,
\end{equation}

\begin{equation}
\label{gleichung61} u_0^{eff} = u_0 - 4 (n+8)\; u_0^2\; L^{-d}
\sum_{{\bf k} \neq {\bf 0}} \left (r_0 - r_{0c} + \hat J_{\bf
k}\right )^{-2}\;.
\end{equation}

Since we work here at finite lattice spacing we have incorporated
the finite shift $r_{0c} = - 4 (n + 2)
u_0 \int\limits_{\bf k} \hat J_{\bf k}^{-1} + O\; (u_0^2)$
of the parameter $r_0$ already at one-loop order.
For simplicity we have not included here the zero-mode average
$M_0^2$ defined previously \cite{esser-1995, chen-dohm-c9-1073-1998} since it
is negligible in the region $T > T_c$. It is convenient to rewrite
$r_0^{eff}$ and $u_0^{eff}$ in terms of the second-moment bulk
correlation length $\xi$. Using the bare one-loop relation
\cite{chen-dohm-b10-1999}

\begin{equation}
\label{gleichung62} r_0 - r_{0c} = J_0 \xi^{-2} \left\{1 + 4 (n
+ 2) u_0 \int\limits_{\bf k} \left [\hat J_{\bf k} \left (\hat
J_{\bf k} + J_0\; \xi^{-2}\right )\right ]^{-1} + O \left (u_0^2
\right )\right\}
\end{equation}

we obtain

\begin{equation}
\label{gleichung63} r_0^{eff} = J_0\; \xi^{-2} - 4 (n + 2)\; u_0\; J_0
^{-1} \Delta_1 \left(\xi^{-2}\right)\;,
\end{equation}

\begin{equation}
\label{gleichung64} u_0^{eff} = u_0 - 4 (n + 8) u_0^2 J_0^{-2}
\int\limits_{\bf k} \left (\xi^{-2} + \hat J_{\bf k} / J_0 \right
)^{-2} + 4 (n + 8)\; u_0^2\; J_0^{-2} \; \Delta_2 \left(\xi^{-2}
\right)\;,
\end{equation}

\begin{equation}
\label{gleichung65} \Delta_m \left(\xi^{-2}\right) =
\int\limits_{\bf k} \left (\xi^{-2} + \hat J_{\bf k} / J_0
\right)^{-m} \;\;- \; L^{-d} \sum_{{\bf k} \neq {\bf 0}} \left
(\xi^{-2} + \hat J_{\bf k} / J_0 \right)^{-m}\;.
\end{equation}

Note that, because of the separation of the zero-mode, the sums in
(\ref{gleichung64}) and (\ref{gleichung65}) do not contain the
${\bf k} = {\bf 0}$ part that was still contained in the function
$\tilde D$, (\ref{gleichung51}). Therefore we obtain the relation
$\Delta_1 \left(\xi^{-2} \right) = L^{-d}\; \xi^2 -
\tilde D (\xi, L, \tilde a)$. The important consequence is that
the power-law term $\propto L^{-d}$
in $\Delta_1 (\xi^{-2})$ becomes dominant for large $L/\xi$ compared
to the exponential behavior $\propto e^{-L/\xi_1}$ of $\tilde D$
according to (\ref{gleichung52}) and (\ref{gleichung43}).
Similarly we have the non-exponential behavior $\Delta_2 \propto
L^{-d}$ for large $L/\xi$.

In the present approximation the susceptibility is

\begin{equation}
\label{gleichung67} \chi = \frac{1}{n} \left(\frac
{L^d}{u_{0}^{eff}}\right) ^{1/2}\; \vartheta_2 \;
\left(Y^{eff}\right)\;,
\end{equation}

\begin{equation}
\label{gleichung68} Y^{eff} = L^{d/2} \; r_0^{eff} \;
\left(u_0^{eff} \right)^{-1/2} \;,
\end{equation}

\begin{equation}
\label{gleichung69} \vartheta_2 (y) = \frac {\int\limits_0^\infty
ds\; s^{n+1}\; \exp\; (- \frac{1}{2}\; y\; s^2 - s^4)}{\int\limits_0^\infty
ds\; s^{n-1}\; \exp\; (- \frac{1}{2}\; y\; s^2 - s^4)}\;.
\end{equation}

These expressions are valid for general $d > 2$ in the sense of
bare perturbation theory. If we take the limit $\tilde a
\rightarrow0$ and apply the field-theoretic RG approach to
(\ref{gleichung67}) - (\ref{gleichung69}) we reproduce the results
of Ref. \cite{esser-1995} for $2 < d < 4$ which are well
applicable to the critical region $0 \leq L/\xi \lesssim O (1)$.
In this region lattice effects are negligible.

In the region $L/\xi \gg 1$ corresponding
to $Y^{eff} \gg 1$ the function $\vartheta_2$ has the
asymptotic expansion \cite{representation}

\begin{equation}
\label{gleichung70} \vartheta_2 (y) = n\; y^{-1} \left [1 - 4 (n+2)
y^{-2} + O \left(y^{-4}\right) \right]\;.
\end{equation}

Within the lowest-mode approximation, where $r_0^{eff} = J_0
\xi^{-2}$ and $u_0^{eff} = u_0$, the expansion (\ref{gleichung70})
implies

\begin{equation}
\label{gleichung71} \chi = J_0^{-1}\; \xi^2 \left [1 - 4(n+2)\; u_0
J_0^{-2}\; \xi^4 / L^d + O \left(L^{-2d}\right)\right ]\;,
\end{equation}

\begin{equation}
\label{gleichung72} \Delta\chi = 4 (n+2) u_0\; J_0^{-2}\; \xi^4\;
L^{-d} + O \left(L^{-2d}\right)\;.
\end{equation}

We see that the lowest-mode approximation yields the incorrect
(non-ex\-po\-nen\-tial) size dependence $\propto L^{-d}$ for large
$L$. In order to see whether this defect is remedied in higher
order we
proceed by including the next terms of (\ref{gleichung63}) and
(\ref{gleichung64}). Then it turns out that the $O (u_0)$
contribution $\propto \Delta_1 \propto L^{-d}$ in (\ref{gleichung63})
cancels the $O (u_0)$ term in (\ref{gleichung71}) and (\ref{gleichung72}).
Therefore $\Delta\chi$ appears to become proportional to $u_0^2$
according to

\begin{equation}
\label{gleichung73} \Delta\chi = -c\; u_0^2 \; J_0^{-4} \; \xi^{8-d}
\; L^{-d} + O \left (L^{-2d} \right ) \;,
\end{equation}
\begin{equation}
\label{gleichung74} c = 8 (n+2) (n+8)\; A_d\; (d-2) (4-d)^{-1}
\;,
\end{equation}
\begin{equation}
\label{gleichung75} A_d = \Gamma (3-d/2) \; 2^{2-d}\; \pi^{-d/2}
(d-2)^{-1}\;,
\end{equation}

for $2 < d < 4$. But in this order the leading size dependence $\Delta\chi \propto
L^{-d}$ in (\ref{gleichung73}) is still non-ex\-po\-nen\-tial. We anticipate that in the next
order of this approach the $u_0^2$ term of (\ref{gleichung73})
will also be cancelled, thus $\Delta\chi \propto u_0^3 L^{-d}$, etc.
We conclude that the present approach does not capture the correct
(exponential) size dependence of $\Delta\chi$ for $L \gg \xi$
{\it{in any finite order}} of perturbation theory. An application of the
renormalization group for $d < 4$ would not remedy this defect for $L \gg
\xi$. In the present context the renormalization of
(\ref{gleichung73}) would only change $J_0^{-2} \; u_0$ to $u^*
\; \xi^{d-4}$ and this would only change the critical
$\it temperature$ dependence of (\ref{gleichung73}) to $\Delta\chi
= - c u^{*2} (L/\xi)^{-d}$ for $d < 4$ in the region $L \gg \xi \gg
\tilde a$.

\end{document}